\documentclass[twocolumn,superscriptaddress,showpacs,preprintnumbers,amsmath,amssymb]{revtex4}
\usepackage{graphicx}
\usepackage{dcolumn}
\usepackage{bm}
\begin{document}
\title{Geometric nature in adiabatic evolution of dark eigenstates}
\author{Shi-Liang Zhu}
\affiliation{FOCUS Center and MCTP, Department of Physics,
University of Michigan, Ann Arbor, MI 48109.}
\author{Z. D. Wang}
\affiliation{Department of Physics and Center of Theoretical and
Computational Physics, University of Hong Kong, Pokfulam Road,
Hong Kong, China}
\begin{abstract}In a recent Letter [Phys. Rev. Lett. {\bf 95},
080502 (2005)], an interesting scheme was proposed to implement a
type of conditional quantum phase gates with built-in
fault-tolerant feature via adiabatic evolution of dark
eigenstates.  In this comment we  elaborate the geometric nature
of the gate scheme and clarify that it still belongs to a class of
conventional geometric quantum computation.\end{abstract}
 \maketitle

In a recent Letter \cite{Zheng}, %
 an interesting scheme was proposed to implement a type of conditional
quantum phase gates with built-in fault-tolerant feature via
adiabatic evolution of dark eigenstates. However, one of the main
conclusions, the proposed conditional phase shift is neither of
dynamic nor geometric origin, is found to be incorrect. Here we
clarify that the phase shift acquired in the proposed gate
operation is a kind of standard geometric phase (GP) defined in
the literature\cite{Aharonov}, and the proposed gate scheme still
belongs to a class of conventional geometric quantum computation
(GQC)\cite{Zanardi}.

In fact, Aharonov and Anandan  showed rigorously that once the
dynamic phase is removed from the total phase shift in any cyclic
evolution of a physical state, the remnant phase must be a
geometric phase connected to a closed curve in the projective
Hilbert space\cite{Aharonov}. So it seems impossible that a phase
shift is neither of dynamic nor geometric origin in any cyclic
evolution. Moreover, a clear geometric picture demonstrated below
enables us to understand the robustness of the proposed gate
scheme.

\begin{figure}[htbp]
\vspace{-0.5cm}
\includegraphics[height=9cm]{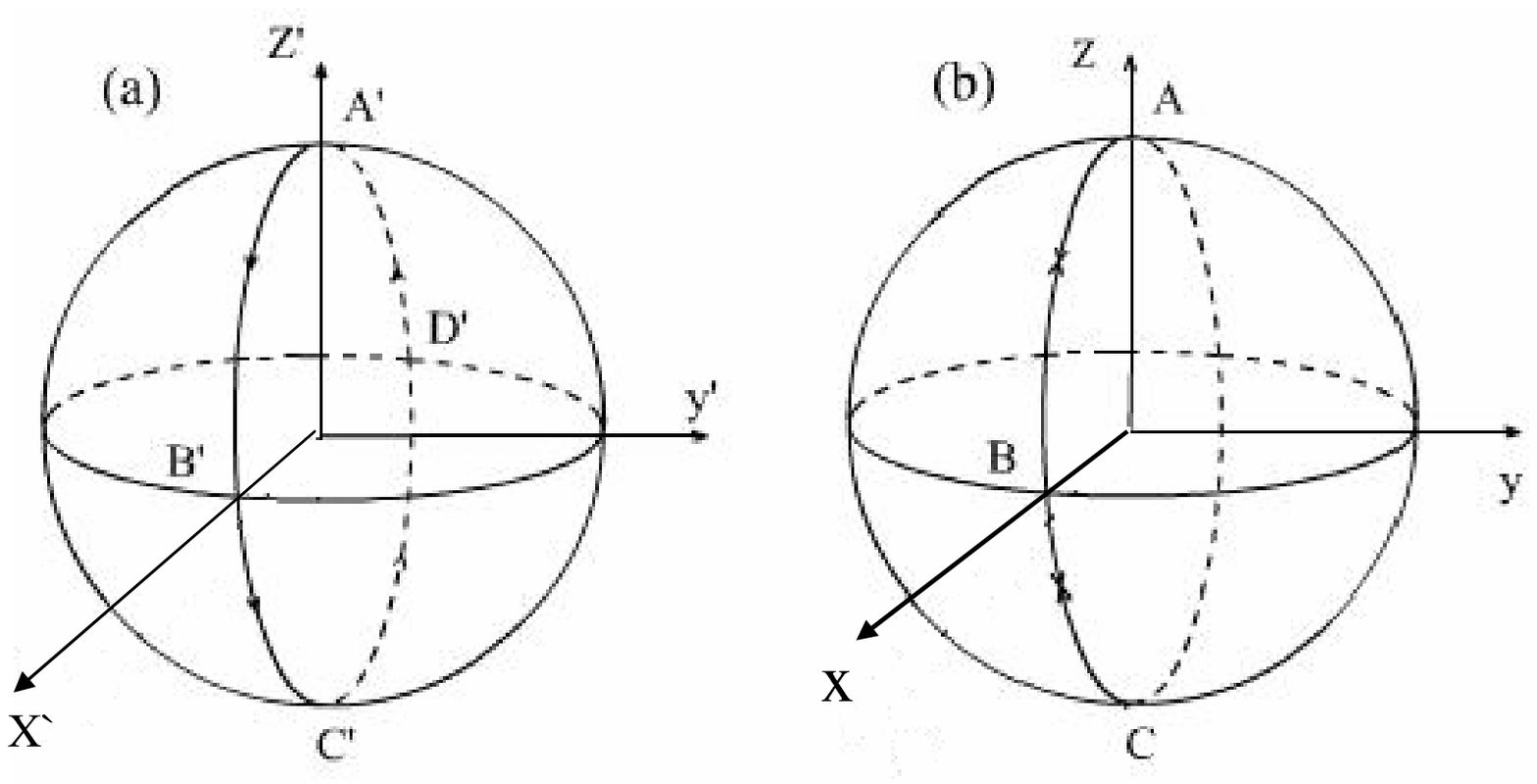}
\vspace{-5.5cm} \caption{The evolution paths in the Bloch sphere.}
\end{figure}

The high-dimensional space in the original system makes its
geometric picture to be implicit. Nevertheless, one can decompose
the whole system into two subspaces. With the same notations as in
Ref. \cite{Zheng}, the Hamiltonian described in Eqs. (1) and (8)
can be truncated in the subspaces $\{
|e_1\rangle|g_2\rangle|0\rangle,
|g_1\rangle|e_2\rangle|0\rangle,|g_1\rangle|g_2\rangle|1\rangle
\}$ and $\{ |e_1\rangle|g_2'\rangle|0\rangle,
|g_1\rangle|e_2'\rangle|0\rangle,|g_1\rangle|g_2'\rangle|1\rangle
\}$, written as

$$
 H_\alpha=\left( \begin{array}{ccc} 0 & 0 & \lambda_1 \\
0 & 0 & -\lambda_2 \\
\lambda_1^\ast & -\lambda_2^\ast & 0
\end{array} \right),\ \   H_\alpha'=\left( \begin{array}{ccc} 0 & 0 & \lambda_1 \\
0 & 0 & c_\alpha\lambda_3 \\
\lambda_1^\ast & c_\alpha\lambda_3^\ast & 0
\end{array} \right),
$$
respectively, where  $c_1=-1$, $ c_2=1$, and $\alpha=1,\ 2$. The
dark state of the Hamiltonian $H_\alpha$ is given by
$|D_\alpha\rangle= (\cos\theta,\sin\theta,0)$, while is
$|D_\alpha'\rangle= (-c_\alpha\cos\theta',\sin\theta',0)$ for the
Hamiltonian $H_\alpha'$. We first analyze the phase shift
accumulated in the evolution of the state $|D_\alpha'\rangle$ that
is actually a two-level state in the subspace
$\{|e_1\rangle|g_2'\rangle|0\rangle,
|g_1\rangle|e_2'\rangle|0\rangle \}$, since the amplitude for the
state $|g_1\rangle|g_2'\rangle|1\rangle$ is always zero. A
standard scenario to look into the geometric structure of  a
two-level system is to study the state evolution on the Bloch
sphere, where any state $|\psi\rangle$ corresponds a point at the
Bloch sphere by the mapping ${\bf n}=\langle \psi |{\bf \sigma}
|\psi \rangle$ with ${\bf \sigma}$ being the Pauli matrix. We now
examine the evolution path evolved in the gate operation proposed
in Ref.\cite{Zheng}: $\theta'$ in $|D_1'(\theta')\rangle$ changes
from $0$ to $\pi/2$ driven by the Hamiltonian $H_1'$, then
$\theta'$ in $|D_2'(\theta')\rangle$ varies from $\pi/2$ to $0$
governed by the Hamiltonian $H_2'$. By a direct calculation, we
have ${\bf n}_\alpha'=(-c_\alpha\sin (2\theta'),0,\cos(2\theta'))$
for the state $|D_\alpha'\rangle$.  The evolution path ${\bf
n}_\alpha'$ of the procedure in the Bloch sphere is plotted in
Fig. 1a. In the first stage, the initial point is $A'$ in Fig.1a,
then the state evolves to the point $C'$ through $B'$. During the
second stage, the state evolves from $C'$ to back $A'$ through
$D'$, and thus
 a closed path in the Bloch sphere is formed. The solid angle
enclosed by the closed path $A'B'C'D'A'$ is evidently $2\pi$, so
the GP acquired is just $\pi$. Similarly,  the evolution path
${\bf n}_\alpha$ determined by the dark state $|D_\alpha\rangle$
is plotted as the path $ABCBA$ with a zero solid angle in Fig. 1b.
Summarizing the results, we illustrate that the GPs
have been acquired in the proposed gate operations described by
Eqs. (10) in Ref.\cite{Zheng}; these GPs are essentially connected
to the phase shifts during the  gate operations.
Therefore,
the proposed gate scheme is still belong to the GQC\cite{Zanardi},
and the above  analysis makes it clear that the robustness of the
quantum gates stems actually  from its geometric nature.


Finally, we wish to remark that the high-dimensional structure
leads to a distinct advantage: the dynamic phases acquired in the
gate operations are automatically zero as the states involved are
dark states, which may simplify the experimental setup. In
contrast, this kind of dark state can hardly be realized in a
single two-level system, and thus an additional operation is
normally required to cancel the dynamic phases\cite{Zanardi}. In
this sense the gate scheme proposed in Ref.\cite{Zheng} is quite
arresting.

The work was supported by the NCET, the NSFC grants (10204008 \&
10429401), and RGC grant of Hong Kong (HKU7114/02P \& 7045/05P)  .

\end{document}